\documentclass[runningheads]{llncs}
\usepackage[T1]{fontenc}
\usepackage{graphicx}
\usepackage[final]{changes}
\usepackage{amsmath,amssymb,amsfonts}
\usepackage{hyperref}
\begin{document}
\title{Triggering Conditions Analysis and \replaced{Use Case}{Methodology} for Validation of ADAS/ADS Functions}
\titlerunning{Triggering Conditions Analysis and Use Case}
\author{V\'ictor~J.~Exp\'osito~Jim\'enez\inst{1} \and
Helmut~Martin\inst{1} \and
Christian~Schwarzl\inst{1} \and
Georg~Macher\inst{2} \and
Eugen~Brenner\inst{2}}
\authorrunning{V. Exp\'osito et al.}
\institute{Virtual Vehicle Research GmbH, Inffeldgasse 21a, 8010 Graz, Austria \\
\email{\{victor.expositojimenez,helmut.martin,christian.schwarzl\}@v2c2.at} \and
Graz University of Technology, Rechbauerstraße 12, 8010, Graz, Austria \\
\email{\{georg.macher,brenner\}@tugraz.at}}
\maketitle              % typeset the header of the contribution
\begin{abstract}
Safety in the automotive domain is a well-known topic, which has been in constant development in the past years. The complexity of new systems that add more advanced components in each function has opened new trends that have to be covered from the safety perspective. In this case, not only the \replaced{specification and requirements}{most common parameters, such as specifications,} have to be covered but also scenarios, which cover all relevant information of the vehicle environment. Many of them are not yet still sufficient defined or considered. In this context, Safety of the Intended Functionality (SOTIF) appears to ensure the system when it might fail because of technological shortcomings or misuses by users. 

An identification of the plausibly insufficiencies of ADAS/ADS functions has to be done to discover the potential triggering conditions that can lead to these unknown scenarios, which might effect a hazardous behaviour. The main goal of this publication is the definition of \replaced{an use case}{ a methodology} to identify these triggering conditions that have been applied to the collision avoidance function implemented in our self-developed mobile Hardware-in-Loop (HiL) platform.

\end{abstract}
%
%
%
% 1 page
\section{Introduction}
\label{section:introduction}

The validation of the Advanced Driver-Assistance Systems (ADAS)/Automated Driving Systems (ADS) has been a topic for numerous research works due to the complexity of functions that have been exponentially growing over the years, since there are more components and software included in each function and their relationships \added{between them (e.g. AI algorithms,...) are getting more challenging }. Reported public cases~\cite{TeslaAccident}~\cite{UberAccident} with human deaths have occurred in the last years, which have given an unreliable \replaced{picture}{image} of the current status of the automated driven cars to the public view. Many standards have been developed to provide a common framework and to overcome these accidents as much a possible in the future in which all aspects related to system dependability have to be covered. The dependability of a function can be split into three main parts according to the origin of the hazard.
Functional Safety hazards, effected by malfunctions based on hardware and \replaced{systematics}{systematically} faults, are covered \replaced{in}{on} the ISO26262~\cite{ISO26262}. Cybersecurity is covered \replaced{in}{on} the ISO/SAE 21434~\cite{SAE21434}, which is focused on external threads.
The ISO21448, Safety of the Intended Functionality (SOTIF)~\cite{ISO21448}, standard is being developed to cover the gap that has been added due to the new requirements and safety standards that ADAS/AD functions developments have to face. With more functions dependent on complex sensors and algorithms, more unknown scenarios could occur in which the function is not designed for. The main goal of the SOTIF standard is to minimize the unknown hazardous scenarios due to technical shortcomings or misuses, specially the ones that can lead to hazardous situations. Many researches~\cite{SecenariobaseExtendedHARAincorporatingFunctionalSafeySOTIFforAD}~\cite{SOTIFProcessAndMethodsInCombinationWithFunctionalSafety} have been trying to reuse and link part of the processes described in the ISO26262 into the SOTIF standard, such as Hazard Analysis and Risk Assessment (HARA) or System-Theoretic Process Analysis (STPA). Another important point for SOTIF is the identification of the limitation of each technology used for automated driving~\cite{IdentificationOfPerformanceLimitationsOfSensingTechnologiesForAutomatedDriving}, which gives more knowledge in order to identify and understand the cause of these unknown scenarios.
To make the steps more explicit for the safety argumentation, the UL4600~\cite{UL4600} provides a checklist of necessary elements to ensure that a function is safe in all aspects. 

The complexity of covering all possible variations and scenarios that can  occur, make validation of ADAS/AD functions a hard task and there are different approaches to face this goal. For example, the validation approach based only on real driven miles data would be unpractical and the cost of the process would be unacceptable~\cite{DrivingSafetyHowManyMilesOfDrivingWouldItTakeDemonstrateAutonomousVehicleReliability}. Another way to support this task would be through simulations. During the last years, new environment simulators specifically designed for this domain, such as CARLA~\cite{CARLA} or SVL Simulator~\cite{rong2020lgsvl}, have been developed. The main issue with this kind of validation is the creation of the realistic high-fidelity sensor models, which are highly complex to develop and also need huge computational resources and high-performance machines~\cite{SOTASensorModelsForVirtualTestingOfAdvanceDriverAssistanceSystemsAutonomousDrivingFunctions} as well as the generation of scenarios~\cite{DevelopmentOfAScenarioSimulationPlatformToSupportAutonomousDrivingVerification} to cover all possible scenario variations. Authors in~\cite{PuttingSafetyOfIntendedFunctionalitySOTIFintoPractice} also include machine learning approaches to get the conclusion that the lack of meaningful trained data makes this approach not ready to fully \replaced{cover}{basis} the whole SOTIF argumentation yet. On the other hand, to avoid the issues previously commented, the authors in~\cite{SelfDrivingCarSafetyQuantificationviaComponentLevelAnalysis} present an approach purely statistical, which could provide a quantitative argument of functions validation. Since some concepts could be reused in our research, in this case, the direction of our research is focused on the decomposition of the function to identify the triggering conditions according to the function design.

\begin{figure}
	\centering
	\includegraphics[width=0.70\textwidth]{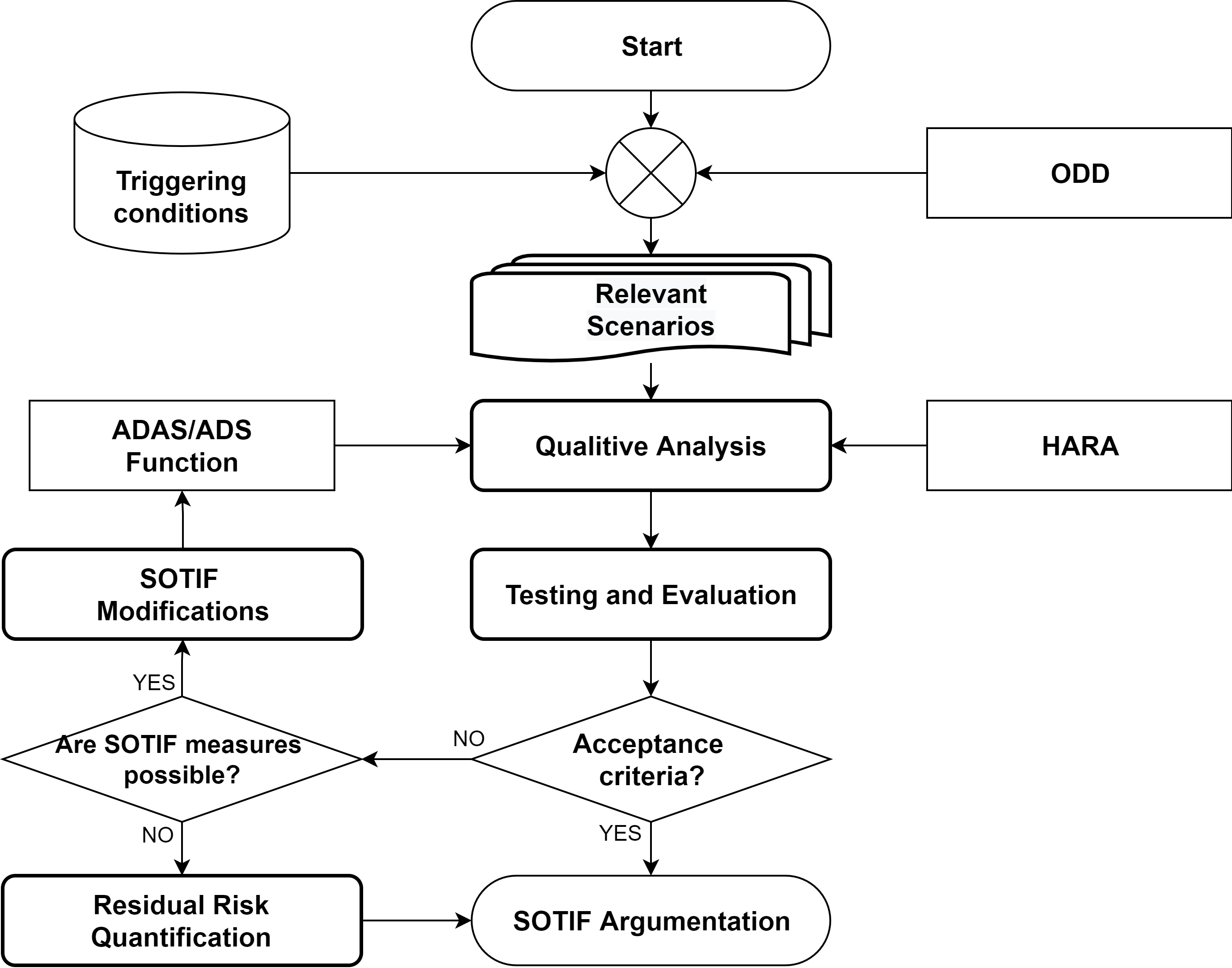}
	\caption{Flowchart of the proposed triggering conditions methodology}
	\label{fig:methodology}
\end{figure}

One of the main step of a SOTIF analysis is the definition of the Operation Design Domain (ODD), according to the standards such as ISO21448 or UL4600. An ODD can be defined as the intended behavior within defined environmental condition that the function has been designed to work on. The definition of the ODD must include all necessary parameters, which make the definition meaningful such as constraints to road conditions. \deleted{For the definition} There are different approaches \added{for the definition of a ODD}, \replaced{from the described in}{from the given in} the PEGASUS project~\cite{9400833} in which the ODD definition is split into six different layers according to the functionality of the parameters to the description given in the SAKURA project~\cite{sakura} or \replaced{the provided by}{the given by} ASAM in ASAM OpenODD~\cite{asamodd}. The authors in~\cite{ODDDescriptionMethodsForAD} present an overview of available approaches for the most common ODD definition. These approaches provide us with the necessary taxonomy to give the basis to create a triggering conditions list. This publication gives an overview of triggering conditions in automated driving~\cite{SafetyTriggerConditionsForCriticalAutonomousSystems}, but in our approach, it is based on scenario level and not the system as it is described. 

The structure of the paper at hand is as follows: \replaced{our approach}{the methodology} is explained in the next section. Section~\ref{section:usecase} goes through all blocks of the proposed \replaced{use case and its implementation in }{and describes them with the help of the selected use case,}the collision avoidance function of our Hardware-in-Loop (HiL) platform. Finally, section~\ref{section:conclusions} shows the conclusions and the future outlook of our research.

\section{Use Case and Methodology}
\label{section:usecase}
The proposed methodology is shown in Figure~\ref{fig:methodology} in which the used blocks and their relationships are depicted. The main goal of this publication is to show the key ideas and process of our \replaced{use case}{methodology} \added{to give the elements to provide a SOTIF argumentation. This is described} as a brief basis and further description will be given in following research work \added{as a more mature methodology.} 

\subsection{ADAS/ADS Function Description}
\label{subsection:function_description}

\begin{figure}
	\centering
	\includegraphics[width=0.50\textwidth]{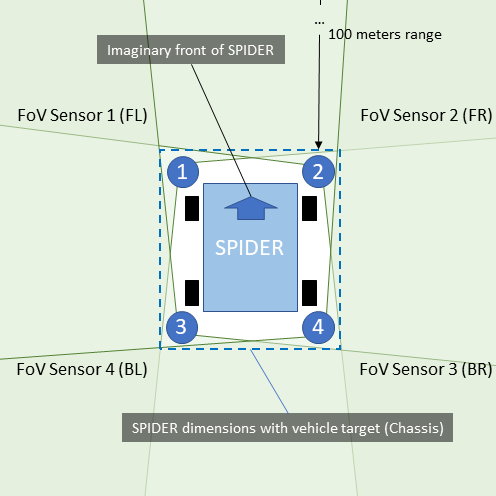}
	\caption{Hardware setup of the collision avoidance function on the HiL Platform (Field of View - FoV, Front Left - FL, Front Right - FR, Back Right - BR, Back Left - BL)}
	\label{fig:caf}
\end{figure}

In this first block, the function and its functionality as well as relevant technical aspects are given in order to be able to know the behaviour of the function. The use case example in this publication is the collision avoidance function implemented in our self-developed mobile Hardware-in-Loop (HiL) platform called SPIDER~\cite{EvaluationOfAnIndoorLocalizationSystemForAMobileRobot}~\cite{spider}. A collision avoidance function could work in two ways, as an Automatic Emergency Brake (AEB) or providing an alternative way/evasive maneuver to reach the target location. For simplicity, only the AEB is implemented at this moment on the robot and it will be used in the analysis in this paper. The function uses four 16-lines lidar sensors located at each corner (Front-Left, Front-Right, Back-Left, Back-Right) to provide redundancy where each point is seen, at least, by two sensors and leave the centre of the robot for different flexible setups in the future as Figure~\ref{fig:caf} shows. The data from the lidars are fused to generate an occupancy grid, which shows the objects surrounding the robot and sends the activation signal to start the emergency brake in case one object \replaced{enters within the delimited zone}{enters with the delimited zone} as the danger zone. According to the SAE J3016 Standard~\cite{SAEJ3016}, it is a Level 4 function since the dynamic driving task fallback has no human interaction.  

To guarantee that there is no collision on the nominal scenario, the Responsibility-Sensitive Safety (RSS)~\cite{shalevshwartz2018formal} model was implemented as the minimum distance before the emergency brake is activated, \replaced{which formula}{which original equation} was adapted for our use case. \added{RSS model was chosen since it provides a proven~\cite{8813885} safe distance according to our robot parameters.} \added{Ego vehicle is defined as the main actor of interest in the scenario, sometimes is also refereed as Vehicle Under Test (VUT). On the other hand, the target vehicle is considered as an element of the scenario and its behaviour is described in the scenario definition.} The assumption of zero speed of the \replaced{vehicle}{velocity} located at front of the ego vehicle\added{(target vehicle)} was set to zero due to the ODD \replaced{is only}{only is} taking \replaced{into}{in} account static objects. Therefore, the RSS equation of the used function is given as follows:

\[
\label{rss_spider}
d_{min} =\left[v_r\rho + \frac{1}{2} a_{max,accel}\rho^2 +
\frac{(v_r + \rho a_{max,accel})^2}{2a_{min,brake}}\right]_+
\]
\[
[x]_+ :=max\{x,0\}.\tag{{2.1}}
\]

\begin{itemize}
	\item\emph{d\textsubscript{min}:} Minimum distance to ensure that there is no crash with the obstacle. 
	\item\emph{v\textsubscript{r}:} Current velocity of the SPIDER (m/s) 
	\item\emph{$\rho$:} Response time in seconds. 
	\item\emph{a\textsubscript{max,accel}:} Maximum acceleration of the robot (m/s\textsuperscript{2}).
	\item\emph{a\textsubscript{max,brake}:} Minimum braking acceleration of the robot~(m/s\textsuperscript{2}). 
\end{itemize}

\subsection{Operational Design Domain}
\label{subsection:odd}

\begin{figure}
	\centering
	\includegraphics[width=0.80\textwidth]{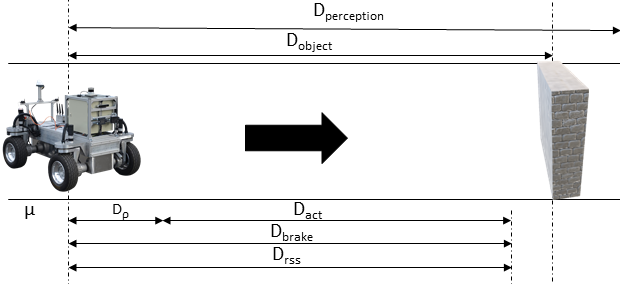}
	\caption{Operational Design Domain (ODD) for the selected use case}
	\label{fig:odd}
\end{figure}

The defined ODD for our use case is a simple one-way road as shown in Figure~\ref{fig:odd} in which there is a static object located at the end of the road within the intended driving path of the robot vehicle. \replaced{In the definition of this ODD, clear weather conditions as well as dry asphalt are assumed, but these conditions would be modified as a part of the test scenarios in the following sections to identify the triggering conditions of the function}{It is also assumed clear weather conditions as well as dry asphalt}. The ego-vehicle, our robot, drives with constant velocity towards the object. 

\begin{itemize}
	\item\emph{D\textsubscript{object}}: Distance between the static object and the robot.
	\item\textit{D\textsubscript{rss}}: Distance calculated by using the RSS equation from~\eqref{rss_spider}.
	\item\textit{D\textsubscript{$\rho$}}: The distance that the robot travels until it perceives the object, makes a decision, and sends an actuation signal.
	\item\textit{D\textsubscript{act}}: The distance between the robot activates the actuation process and it is finished.
	\item\textit{D\textsubscript{brake}} the distance necessary to brake the robot until velocity is zero. $ D\textsubscript{brake} = D\textsubscript{act} + D\textsubscript{$\rho$}$. In nominal situations, $ D\textsubscript{brake} = \textit{D\textsubscript{rss}}$.  
	\item\textit{D\textsubscript{perception}}: the perception range of the sensors. \textit{D\textsubscript{perception}} should be longer than \emph{D\textsubscript{object}} and \textit{D\textsubscript{brake}}.
	\item\textit{$\mu$}: the friction factor of the road. 
\end{itemize}

According to the technical parameters of the function (\emph{v\textsubscript{r}}:~50km/h, $\rho$:~1~second, \emph{a\textsubscript{max,accel}}:~2.0m/s\textsuperscript{2}, and \emph{a\textsubscript{max,brake}}:~5.0m/s\textsuperscript{2}), and the used RSS model, is assumed that the maximum RSS minimum distance (\emph{D\textsubscript{rss}}) will always be smaller than the range of the perception sensors ($D\textsubscript{rss,max}~\textless~D\textsubscript{perception}$). 

%Once the ODD is defined, the specific scenario included the  
%- What is a ODD?
%- Definitions
%- Examples, ALKS, Highway Pilot
% We could include research from MASA --> use as less information as possible from MASA

\subsection{Triggering Conditions List}
\label{subsection:triggering_conditions}
According to the ISO21448~\cite{ISO21448} the triggering conditions are the specific conditions of a scenario that serve as an initiator for a subsequent system reaction leading to hazardous behaviour. To identify the possible conditions that can bring up the function to unknown hazards, \replaced{parameters of the defined ODD will be modified}{all possible parameters of the ODD have to be proven}. \replaced{An own database with the potential triggering conditions based on the taxonomy described in Annex C of the ISO21448\cite{ISO21448} and the BSI standard~\cite{BSI} was developed for this purpose. This database includes categories and different levels of subcategories, which adds granularity to each subcategory. For example, the main category is at the top (e.g. environmental conditions, road conditions,...) and the more granular conditions are at the bottom (e.g. heavy snow). The main category, environmental conditions in this example, follows other subcategories such as the \textit{weather conditions} or \textit{illumination} and, then, these subcategories are split again into other subcategories. In the case of \textit{weather conditions}, this subcategory is split into rain, snow, fog and other climate conditions to continue with more fined granularity, which includes different intensities as light, medium and heavy.}{The triggering conditions follows a hierarchy system from top to down in which the main category is at the top (e.g. environmental conditions) and the most granular conditions are a the bottom (e.g. heavy snow). For example, environmental conditions follows to the weather conditions and, then, they are split in rain, snow, fog and other climate conditions to continue with more fined granularity, which includes different intensities as light, medium and heavy. The taxonomy used of these triggering conditions list is a combination of the proposal in Annex C of the ISO21448\cite{ISO21448} and the BSI standard~\cite{BSI}}. The main \replaced{purpose}{goal} of using this \replaced{approach is to provide granularity}{granularity is} to cover the maximum variety of situations, which should help to discover more shortcomings of function during the analysis.

\subsection{Relevant Scenarios}
\label{subsection:scenario}

\replaced{In our use case, a scenario is defined as the combination of the previously defined ODD combined with potential triggering conditions. Based on this definition, the relevant scenarios are the compilation of scenarios that includes the selected triggering conditions in which the output of the function will be analysed to know whether the selected triggering condition could lead to a hazardous behaviour}{The compilation of the scenarios is done by combination of the defined ODD with the selected triggering condition that operates the function in the specific environmental condition to analyse whether it can lead to a hazardous behaviour}. \replaced{In this step, only the triggering conditions that makes senses for the defined ODD will be taken into account}{Another point to take into account is that the triggering conditions list has to be filtered to make sense with the selected ODD. For example, in this use case, the triggering conditions related to another target vehicle are omitted since they are not relevant for the defined ODD in the selected use case}. For simplicity, in a first iteration only a single triggering condition is added to the ODD to better identify the cause and origin of the hazard.

%\begin{figure*}
%	\centering
%	\includegraphics[width=1.00\textwidth]{resources/CAF_architecture_v06.png}
%	\caption{Block decomposition of the collision avoidance function}
%	\label{fig:limitations}
%\end{figure*}

\subsection{Hazard Analysis and Risk Assessment}
\label{subsection:hara}
HARA methodology is always performed in Functional Safety processes in which associated hazards to the system are analysed and classified according to the level of danger and caused injuries. This classification used the Automotive Safety Integrity Level (ASIL) which uses the parameters, Exposure (E), which is the possibility of occurrence of the hazard within the operation situation; Severity (S), which is the estimation of the harm that the hazard might cause; and Controllability (C) that is ability to avoid the harm (see ISO262626~\cite{ISO26262} part 3). But unlike the process implemented in fuctional safety standard, the ASIL level is not used in the SOTIF analysis process and only the identification of the hazards are used. 

As a part of the HARA analysis and the given scenario, the following identified hazards can be extracted: \textit{"a collision due to the inability to completely brake before reaching an obstacle"} and \textit{"activation of the emergency brake due to a false object detection"} 

\subsection{Qualitative Analysis}
\label{subsection:analysis}

\begin{figure*}
	\centering
	\includegraphics[width=1.0\textwidth]{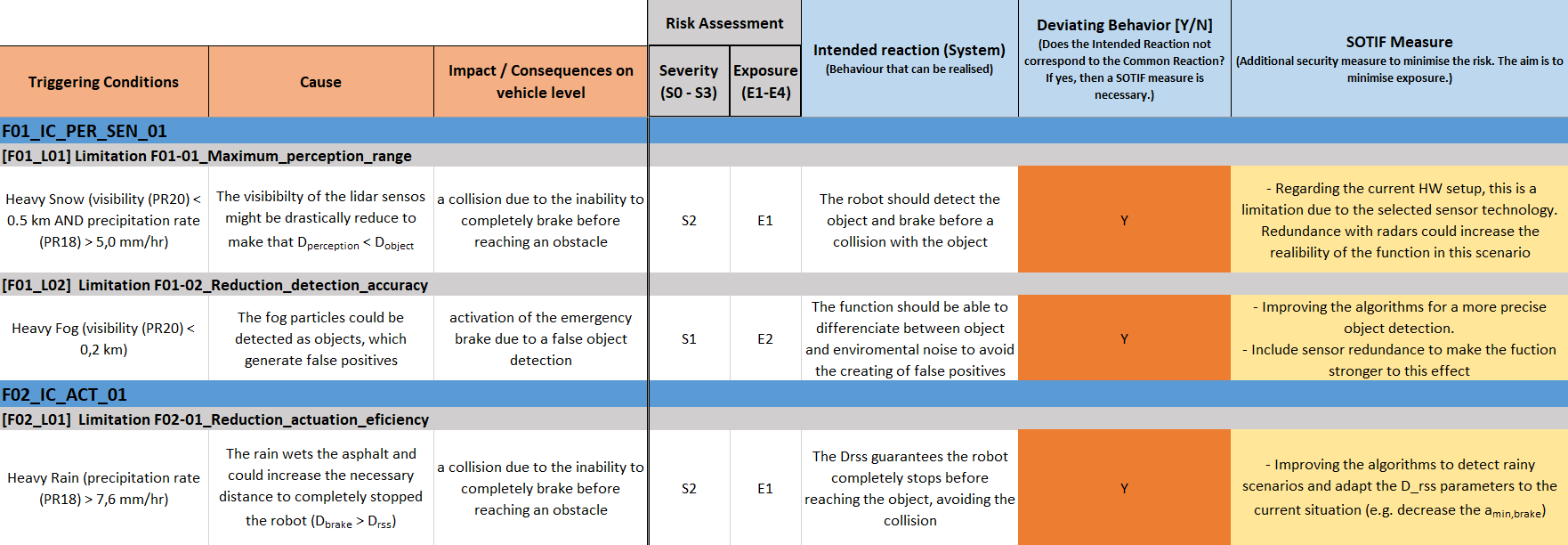}
	\caption{Triggering Conditions Analysis Sheet Excerpt}
	\label{fig:tca}
\end{figure*}

In this step, the impact of the defined scenarios on the function has to be analysed to know if they could lead to a performance limitation which could further effect a hazard. To do this, first, the function is split into the three main categories: perception, decision, and actuation. \deleted{Figure~\ref{fig:limitations} depicted this decomposition for the proposed use case.} The perception part is also divided into perception sense and perception algo\added{, a shorter reference to perception algorithms}. This differentiation is done to separate the part in which perception only provides data from the perceived reality through sensor input and perception algo in which information is extracted based on  that perceived data to get any environmental information, e.g. objects, driving road. In this use case, the perception sense includes the sensor drivers, data fusion, and some data post-processes like filtering ground from the point clouds. The block perception algo includes the costmap generator, which generates an occupancy grid from the filtered point cloud to determine the objects that surround the SPIDER. Decision part includes the node, called collision detector, that goes through the generated occupancy grid and detects the closest object and its distance to the robot. The output of the node is the trigger of the emergency brake in case the object violates the calculated allowed minimum distance or the object distance if an object is detected but the emergency brake is not yet necessary. Finally, the actuation part is done by the commands to steering, braking and propulsion. In the given use case it includes the emergency brake from the SPIDER. The main reason for this categorization is to identify the components and their location in the system architecture in which the performance limitations could \replaced{occur}{happen}. For this reason, collaboration with domain experts and developers is crucial to better identify the origin of the hazards. 
A discussion has to be done for each selected scenario to identify how it could affect the function output. For example, for the scenario that includes medium or high snow level, these specific conditions may affect the perception sense part of the function, since it can include ghost points in the perceived data, and the severity of the effect of the limitation is further rated by NO/Low/Medium/High (S0-S3). Therefore, this triggering condition could lead to the previously identified hazard \textit{"activation of the emergency brake due to false object detection"}. Another specific scenario is the one in which the surface is slippery due to heavy rain. This kind of surface reduces the friction of the wheels (actuation part) and increases the distance necessary to completely stop the robot. This triggering condition has an impact on the surface friction of the ODD and could lead to the hazard \textit{"a collision due to the inability to completely brake before reaching an obstacle"}. Similar discussions have to be done through the whole list of selected scenarios to identify the potential hazardous scenarios. An excerpt of this analysis can be shown in Figure~\ref{fig:tca}. In this use case, it was assumed the controllability (C) of the function is always C3 since it is a level 4 function and no backup operator is implemented. 
% If a message like this is received: LaTeX Error: Cannot determine size of graphic in resources/tca_v05.png (no BoundingBox). Use the followig line:
% It also possible to use this: \includegraphics[scale=1,bb=0 0 30 30]

\subsection{Testing and Evaluation}
\label{subsection:acceptance_criteria}

The first step in this process is the definition of the Key Performance Indicators (KPIs)\added{,} which captures the \replaced{behaviour}{performance} in a nominal scenario and \replaced{set the considered acceptable criteria}{set the criteria that is considered acceptable} to release the function. This nominal scenario, sometimes also called functional scenario, is the one used in optimal conditions without the inclusion of any of the triggering conditions. An example of a KPI could be Time To Collision (TTC)~\cite{TTC}, which represents the time that the ego vehicle needs until it reaches the target  vehicle or an obstacle according to the current ego and target \replaced{velocities}{vehicle} and its distance between \replaced{the two vehicles}{them}. \added{This metric can be defined by the formula~\eqref{ttc_formula}. Similar to RSS formula~\eqref{rss_spider}, it is assumed that $v\textsubscript{target} = 0$ since there is a static object instead of a target vehicle defined in the ODD.}

\[
\label{ttc_formula}
TTC = d / (v\textsubscript{ego}-v\textsubscript{target}).\tag{2.2}
\]

Afterwards, the results obtained in the different levels of the verification and validation procedures \replaced{have}{has} to be compared with the nominal conditions and ensure they are within the defined acceptable limits. 

\subsection{SOTIF Modifications}
\label{subsection:sotif_modifications}

The main goal of applying SOTIF measures is to improve the output of the function and minimise the risk level that applied triggering conditions might lead to. Following the goal of the the principle of ALARP (As Low As Reasonably Practicable)~\cite{MakingALARPDecisionSufficientTesting}, which assumes \replaced{that the efforts to achieve zero risk could be not feasible}{that zero risk is not possible} and it has to be \added{taken} into account the necessary efforts needed to reduce the risk \added{as much as reasonable}. For example, there are some cases in which it is not possible to further risk reduction or the associated efforts to reduce it are not worth it. Returning to the proposed use case, in that case where a weather conditions is included as potential triggering conditions in the scenario as heavy snow, the visibility is drastically reduced \deleted{(in extreme situations where $\textit{D\textsubscript{perception}~\textless~D\textsubscript{object}}$), according to the BSI 1883:2020~\cite{BSI} standard}. There is a chance that the snow could limit the perception range of the sensors and it can lead to the previously defined hazard: \textit{"a collision due to the inability to completely brake before reaching an obstacle"}\added{ in which the relationship $\textit{D\textsubscript{perception}~\textless~D\textsubscript{brake}}$ could apply to lead to the specified hazard}. \added{According to the situation, modifications have to be applied to improve the output on the specified scenario. For example, diversity of sensor technologies, which are stronger against the described issue or an external hardware to measure the specific triggering condition (fog level on the environment) could be applied to mitigate the occurrence of the hazard. }In cases in which the effectiveness of the SOTIF measures are not enough to pass the defined acceptance criteria level, these scenarios have to be considered and evaluated within the \textit{"Residual Risk Evaluation"} block.

\subsection{Residual Risk Quantification}
\label{subsection:risk_evaluation}

In case the maximum efforts to minimize the risks are reached \added{and there is not other possible way or feasible way to further improve the function}, the evaluation of the existing residual risk has to be calculated. The risk could be expressed in a function as follows:  

\[R = f(S, O)\]

According to the given formula, the risk~(\(R\)) can be defined by the severity~(\(S\)) and probability of occurrence (\(O\)). In this function, the severity is the specified in the analysis for the associated hazard. On the other hand, finding the right occurrence probabilities of each triggering condition presents a more complex problem since there are a wide variety of situations from weather to road conditions. Moreover, finding or calculating these probabilities could change between countries, where the availability of some of them may not be publicly accessible, which makes the problem even harder. At the end of this block, a quantitative evaluation of each identified risk (\(R\textsubscript{0}... R\textsubscript{N}\)) should be provided. For example, the number of kilometres or the number of hours until the function could lead to a hazard.

\subsection{\replaced{SOTIF Argumentation}{End of the Argumentation}}
\label{subsection:end_argumentation}

Finally, the output of the previous blocks, which includes the carried out measures covered by functional modification and verification measures, such as simulations and testing, as well as the remained evaluated risks for each triggering condition, are included as inputs for the results of the SOTIF argumentation where both, reports (e.g. identified triggering conditions, relevant scenarios...) and metrics (e.g. number of hours until hazards behavior appears\added{, KPIs,...}), should be included. 

\section{Outlook}
\label{section:conclusions}

In this publication, a methodology to extract and evaluate the triggering conditions of an ADAS/ADS function is described. Since driven kilometres or simulations may not be enough in some situations, by using the introduced \replaced{approach}{methodology}, a profound understanding of the system as well as the discovery of unknown hazardous scenarios could be provided. The addition of the triggering conditions analysis in the\replaced{use case}{methodology} helps the engineers to discover the ones that might lead to a hazard, and also provide more parameters in order to evaluate the risks of the function. For this purpose, the paper goes through all blocks of the proposed \replaced{approach}{methodology} in which a description and a goal for each one is given as well as the relationship between them. The \replaced{use case}{methodology} is illustrated by using the collision avoidance function implementation in our mobile \added{robot} HiL platform \deleted{robot vehicle use case}.

In future research works, the residual risk evaluation will be investigated in more detail to be able to provide a quantitative argument at the end of the process, which can provide specific metrics (e.g. validated number of kilometres) that will be used to characterize the safety of a function to achieve the demanded acceptance criteria. Moreover, in the following research, we want to \replaced{extend}{use} this methodology in different approaches of the collision avoidance function to see how different sensor setups (e.g., RADAR+LIDAR, RADAR+LIDAR+CAMERA) can affect the function validation by combination of simulation and real-world testing. 

\section*{Acknowledgments}

Research leading to these results has been performed in the project FRACTAL A Cognitive Fractal and Secure EDGE based on an unique Open-Safe-Reliable-Low Power Hardware Platform Node, under grant agreement No 877056. The project is co funded by grants from Germany, Austria, Finland, France, Norway, Latvia, Belgium, Italy, Switzerland, and Czech Republic and -Electronic Component Systems for European Leadership Joint Undertaking (ECSEL JU). The publication was written at Virtual Vehicle Research GmbH in Graz and partially funded within the COMET K2 Competence Centers for Excellent Technologies from the Austrian Federal Ministry for Climate Action (BMK), the Austrian Federal Ministry for Digital and Economic Affairs (BMDW), the Province of Styria (Dept. 12) and the Styrian Business Promotion Agency (SFG).

\bibliographystyle{IEEEtran}
\bibliography{bibliography}
\end{document}